\documentclass[final]{jfm}
\usepackage{graphicx,xcolor,amssymb,amsmath}
\usepackage{dcolumn}% Align table columns on decimal point
\usepackage{bm}% bold math
\usepackage{hyperref}
\usepackage{natbib}
\usepackage[T1]{fontenc}

\ifCUPmtlplainloaded \else
  \checkfont{eurm10}
  \iffontfound
    \IfFileExists{upmath.sty}
      {\typeout{^^JFound AMS Euler Roman fonts on the system,
                   using the 'upmath' package.^^J}%
       \usepackage{upmath}}
      {\typeout{^^JFound AMS Euler Roman fonts on the system, but you
                   dont seem to have the}%
       \typeout{'upmath' package installed. JFM.cls can take advantage
                 of these fonts,^^Jif you use 'upmath' package.^^J}%
      }
  \else
  \fi
\fi

\ifCUPmtlplainloaded \else
  \checkfont{msam10}
  \iffontfound
    \IfFileExists{amssymb.sty}
      {\typeout{^^JFound AMS Symbol fonts on the system, using the
                'amssymb' package.^^J}%
       \usepackage{amssymb}%

      }{}
  \fi
\fi

\ifCUPmtlplainloaded \else
  \IfFileExists{amsbsy.sty}
    {\typeout{^^JFound the 'amsbsy' package on the system, using it.^^J}%
     \usepackage{amsbsy}}
    {\providecommand\boldsymbol[1]{\mbox{\boldmath $##1$}}}
\fi

\newcommand{\rmi}{\mathrm{i}}
\newcommand{\rmd}{\mathrm{d}}

\newcommand{\bx}{\boldsymbol{x}}

\newcommand{\bX}{\boldsymbol{X}}

\newcommand{\bv}{\boldsymbol{v}}

\newcommand{\bs}{\boldsymbol{s}}
\newcommand{\bsd}{\dot{\boldsymbol{s}}}
\newcommand{\st}{\varsigma}

\newcommand{\hatx}{\mathbf{\hat{x}}}
\newcommand{\haty}{\mathbf{\hat{y}}}
\newcommand{\hatz}{\mathbf{\hat{z}}}
\newcommand{\hatn}{\mathbf{\hat{n}}}
\newcommand{\hatN}{\mathbf{\hat{N}}}
\newcommand{\hatt}{\mathbf{\hat{t}}}

\newcommand{\bN}{\mathbf{N}}
\newcommand{\bG}{\mathbf{P}}
\newcommand{\bS}{\mathbf{S}}

\newcommand{\bF}{\mathbf{f}}
\newcommand{\bp}{\mathbf{p}}

\newcommand{\phialt}{{\tilde\phi}}

\newcommand{\ce}[1]{{\color{black}#1}}

\title{Flow associated with Lighthill's elongated-body theory} 
\shortauthor{C. Eloy, S. Michelin} 

\author{
	Christophe Eloy\aff{1}\corresp{\email{celoy@centrale-med.fr}}
	\and
	S\'ebastien Michelin\aff{2} 
	}

\affiliation{
	\aff{1}Aix Marseille Univ, CNRS, Centrale Med, IRPHE, Marseille, France \\
	\aff{2}LadHyX, CNRS - Ecole Polytechnique, Institut Polytechnique de Paris, 91120 Palaiseau, France
}

\begin{document}

\maketitle

\begin{abstract} 
The hydrodynamic forces acting on an undulating swimming fish consist of two components: a drag-based resistive force and a reactive force originating from the necessary acceleration of an added mass of water. Lighthill’s elongated-body theory, based on potential flow, provides a framework for calculating this reactive force. By leveraging the high aspect ratio of most fish, the theory simplifies the problem into a series of independent two-dimensional slices of fluids along the fish's body, which exchange momentum with the body and neighbouring slices. Using momentum conservation arguments, Lighthill's theory predicts the total thrust generated by an undulating fish, based solely on the dimensions and kinematics of its caudal fin. However, the assumption of independent slices has led to the common misconception that the flow produced lacks a longitudinal component. In this paper, we revisit Lighthill’s theory, offering a modern reinterpretation using essential singularities of potential flows. We then extend it to predict the full three-dimensional flow field induced by the fish’s body motion. Our results compare favourably with numerical simulations of realistic fish geometries.
\end{abstract}

%%%%%%%%%%%%%%%%%%%%%%%%%%%%%%%%%%%%%%%%%%%%%%%%%%%%%%%%%%%%%%%%%%%%%%%
\section{Introduction}

Most fish and cetaceans propel themselves by undulating their backbone, a mode of locomotion known as undulatory swimming \citep{Lauder2005}. 
At low Reynolds numbers, the dynamics and energetics of undulatory swimming can be  described theoretically using \emph{resistive} forces, which depend on the relative velocity between the body and the fluid \citep{Taylor1952,Gray1955}. 
At high Reynolds numbers, new forces arise: \emph{reactive} forces. These reactive forces are related to the acceleration of an added mass of water due to a force that the body applies on water and that, reactively, the water applies to the body \citep{Lighthill1971}. 
To estimate those reactive forces, the natural framework is potential flow theory. 

Potential flow theory assumes that the flow is inviscid and irrotational almost everywhere, i.e. that vorticity can be located in singular lines or sheets attached to the boundaries and possibly shed in the wake. 
Under such a hypothesis, the flow velocity is the gradient of a potential, which is a solution of a Laplace equation through the incompressibility condition. 
This Laplace equation is solved by enforcing the Neumann boundary conditions associated with the impermeability of the body (hence, the normal flow velocity is imposed on the body). 
In the 1960s and 1970s, unsteady potential flow theory was used to describe undulatory swimming in two limits: the two-dimensional limit describing the caudal fin as an oscillating two-dimensional foil \cite{Wu1961,Wu1971a}, and the slender-body limit considering small cross-section compared to the body length \cite{Lighthill1960}.
 
Slender-body theory was proposed by \cite{Munk1924} as a method for analyzing the aerodynamic forces acting on airship hulls. \cite{Munk1924} explored the dynamics of a rigid body moving through a stationary fluid. 
This theory relies on the conservation of fluid momentum along the body's axis. 
The key idea is to exploit the asymptotically large aspect ratio to assume that the flow induced by each section does not significantly influence distant sections. 

\cite{Lighthill1960} further expanded slender-body theory to compute the fluid dynamics around a flexible body in motion. 
He demonstrated that, in the case of a periodically deforming fish, the average thrust force only depends on the tail's kinematics. 
However, this theory is limited to displacement amplitudes that are small compared to the body length. To address this limitation, \cite{Lighthill1971} later expanded slender-body theory, introducing what is now known as \emph{Lighthill's elongated-body theory}. The goal of the present study is to revisit this seminal article to clarify some misunderstandings and show that Lighthill's elongated-body theory is associated with a flow that can be easily computed and compared to direct numerical simulations. 

Lighthill's elongated-body theory has significantly deepened our understanding of swimming dynamics and energetics. To date, this is the only theoretical approach that allows one to evaluate the reactive forces applied on a swimming body. Although some researchers criticised Lighthill's theory, arguing that these potential-based approaches tend to overestimate forces \citep{Hess1984,Anderson2001,Shirgaonkar2009}, it has been successfully used to study canonical fluid-structure instabilities \citep{michelin2008vortex,buchak2010clapping}, to assess whether fish employ smart drag reduction mechanisms \citep{Alexander1977,Webb1975,Videler1981,Ehrenstein2013}, to design controllers for swimming robots \citep{Boyer2010,Porez2014}, to perform optimisation calculations \citep{eloy2012optimal,eloy2013best}, and to study energy harvesting devices \citep{Singh2012,Michelin2013}. 
\cite{Candelier2011} extended Lighthill's elongated-body theory to account for arbitrary three-dimensional motions, including rotations, and \cite{Candelier2013} introduced a method to compute reactive forces in a non-uniform background flow.

\ce{The goal of this article is to calculate Lighthill's reactive force with an alternative method and compute the resulting flow.} 
Before doing so, \S\ref{sec:straight} will first introduce the problem through the canonical configuration of a rigid and straight cylinder moving in an inviscid fluid. In \S\ref{sec.Alternative_derivations}, we will then use two distinct methods to calculate Lighthill's reactive force: one based on momentum conservation similar to the original derivation of \cite{Lighthill1971}, and one based on the reconstruction of the flow with a distribution of dipoles and sources. Then, in \S\ref{sec.Comparison_numerics}, we apply this framework to compare the flow resulting from Lighthill's elongated-body theory to numerical simulations before discussing our results in \S\ref{sec.Discussion}.

%%%%%%%%%%%%%%%%%%%%%%%%%%%%%%%%%%%%
\section{Flow around a straight cylinder}
\label{sec:straight}

%%%%%%%%%%%%%%
\begin{figure}
\begin{center}
\includegraphics[width=0.713\textwidth]{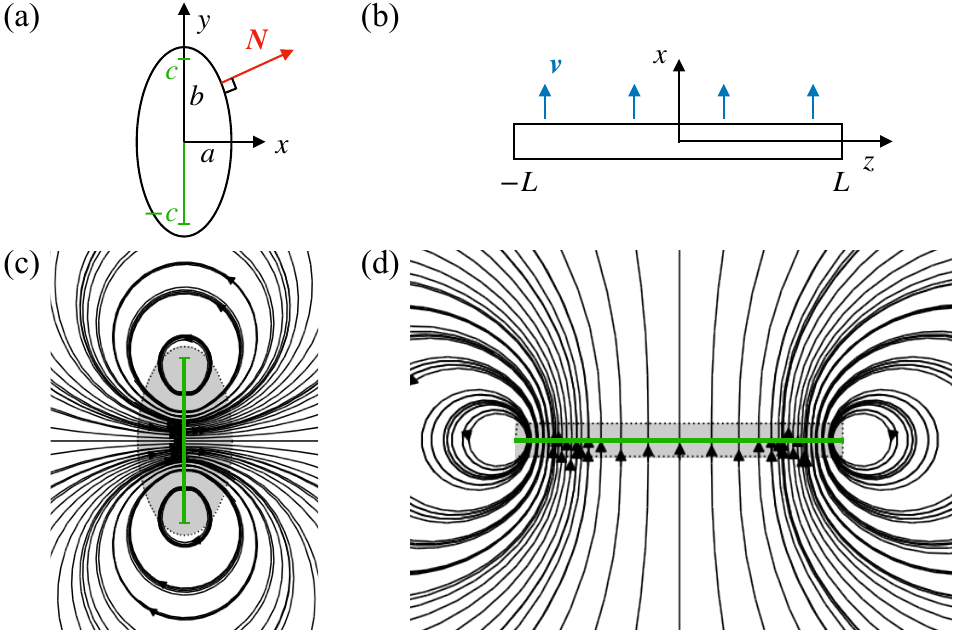}
\caption{
(a) We consider a straight cylinder of length $2L$ with elliptic cross-section with semi-minor axis $a$ and semi-major axis $b$. 
(b) This cylinder is moving with a velocity $v(t)$ along the $x$ direction.
(c-d) The streamlines of the potential flow are obtained by integrating the dipoles distributed along a ribbon spanning between the foci of the ellipse (represented in green).
}
\label{fig:straight}
\end{center}
\end{figure}
%%%%%%%%%%%%%%

Before examining the case of a swimming fish, it is worth first considering a straight cylinder of finite length $2L$ and elliptic cross-section with semi-minor axis $a$ and semi-major axis $b$. This straight cylinder moves perpendicularly to its axis with a velocity $\bv = v\, \hatx$ in a perfect fluid at rest (figure~\ref{fig:straight}). 
To reconstruct the flow, we use a surface distribution of dipoles distributed along a ribbon spanning between the elliptical section's foci of coordinates $\eta=\pm c$ along the major axis, with $c=\sqrt{b^2-a^2}$ the linear eccentricity of the ellipse. 
For an infinitely long cylinder, the distribution of dipolar intensity is given by $\bG(\eta) = P(\eta)\,\hatx$ with 
\begin{equation}\label{eq:P_straight}
P(\eta) = \frac{2bv}{b-a} \sqrt{c^2 - \eta^2},
\end{equation}
and $-c<\eta<c$. 

For a finite cylinder of large aspect ratio ($L\gg b$), we will assume that the distribution is the same away from the cylinder's ends. 
The associated potential is given by
\begin{subequations}
\begin{eqnarray}
\phi(\bx) & = & \int_{-L}^{L} \int_{-c}^{c} -\frac{\bG(\eta)\cdot(\bx-\bX)}{4\pi\|\bx-\bX\|^3}\rmd \eta \rmd s,\\
		  & = & \int_{-c}^{c} -\frac{P(\eta)x}{2\pi (x^2 + (y-\eta)^2)} g(z, x^2 + (y-\eta)^2 , L) \rmd \eta, \label{eq:phi_straight}
\end{eqnarray}
\end{subequations}
where $\bX = (0,\eta,s)$ is a point along the ribbon, $\bx = (x,y,z)$ is a point in the fluid domain, and $g$ is a correction due to the finite length of the cylinder with 
\begin{equation}
g(z,r^2,L) = 
		\frac{L+z}{2\sqrt{(L+z)^2+r^2}} + \frac{L-z}{2\sqrt{(L-z)^2+r^2}}\cdot
\end{equation}
When $\bx$ is located close to the cylinder (when compared to its length), $r\ll L$, $g\sim 1$ and we recover the bidimensional flow around an infinite cylinder; when $\bx$ is far from the cylinder, $r\gg L$, $g\sim L/r$ and the flow corresponds to a three-dimensional dipole. 

Far from the cylinder's ends, and in the vicinity of the cylinder's surface (i.e. $r\ll L$), the potential $\phi$ is given by \eqref{eq:phi_straight} with $g\approx 1$. This potential satisfies the following boundary condition of impermeability on the cylinder's surface  
\begin{equation}\label{eq:BC_straight}
\bnabla \phi \cdot \bN = \bv \cdot \bN, 
\end{equation}
with $\bN = b \cos \theta\, \hatx + a \sin \theta\, \haty$ a (non-unit) vector normal to the body's surface. The proof is given in appendix \ref{sec.appendix1} using the formalism of complex potentials. 

The pressure field is calculated from the generalised Bernoulli equation, 
\begin{equation}\label{eq:p_straight}
p = p_\infty - \frac{1}{2}\rho \|\bnabla \phi\|^2 - \rho \dot{\phi} +  \rho \,\bv \cdot \bnabla\phi,
\end{equation}
with $\rho$ the fluid density, \ce{$p_\infty$ the pressure at infinity and the last term $\rho \,\bv \cdot \bnabla\phi$ arises because the reference frame is a moving frame attached to the cylinder.}

From \eqref{eq:phi_straight} and \eqref{eq:p_straight},  the force per unit length exerted by the fluid onto the cylinder is obtained as
\begin{equation}\label{eq:f_straight}
\bF = \int_{\cal{C}} - p \hatN \rmd \ell
	= \int_{-\pi}^{\pi} - p\, \bN\,\rmd \theta 
    = -m_a\dot{v}\, \hatx,
\end{equation}
where $\cal{C}$ is the contour of the ellipse, $\rmd\ell$ its elementary arclength, $\hatN$ its unit normal vector, and $m_a$ is the added mass per unit length of the cylinder,
\begin{equation}\label{eq:added_mass}
 m_a = \rho \pi b^2.
\end{equation}
The details of this calculation are given in appendix~\ref{sec.appendix2}. 
Note that only the term $\rho \dot{\phi}$ of~\eqref{eq:p_straight} contributes to the integral in~\eqref{eq:f_straight}, because it is the only $x$-odd component. 

The added mass $m_a$ can be used to calculate other useful quantities. For instance, in a plane perpendicular to $\hatz$, the fluid's kinetic energy and momentum (per unit axial length $z$) are $E_c= \frac{1}{2} m_a v^2$ and $\bp = m_a v \hatx$, respectively, while the total axial pressure force that the upstream fluid ($z<0$) \ce{exerts on the downstream fluid ($z>0$) is $\bF_p = \frac{1}{2} m_a v^2 \hatz$ (figure~\ref{fig:notations}c).}

%%%%%%%%%%%%%%%%%%%%%%%%%%%%%%%%%%%%%%%%%%%%%%%%%%%%%%%%%%%%%%%%%%%%%%%
\section{Alternative derivations of Lighthill's elongated-body theory}
\label{sec.Alternative_derivations}

We now present two alternative derivations of Lighthill's reactive force on an elongated body, starting in \S\ref{sec:momentum_derivation} with Lighthill's original approach based on momentum conservation adapted to a slice of fluid. We then show in \S\ref{sec:dipole} how to obtain the same force by representing the flow potential with a distribution of sources and dipoles.

%%%%%%%%%%%%%%%%%%%%%%%%%%%%%%%%%%%%
\subsection{Problem statement}
\begin{figure}
\begin{center}
\includegraphics[width=.99\textwidth]{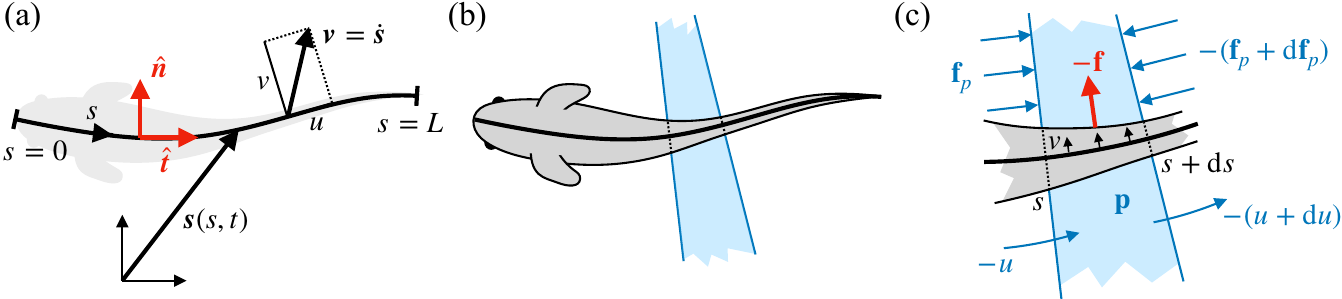}
\caption{
(a)	An elongated fish is swimming by undulating its backbone. The vector $\boldsymbol{s}(s,t)$ describes the planar motion of the backbone.
(b--c) Illustration of the argument of momentum conservation, originally proposed by \cite{Lighthill1971}, and here adapted to a slice of fluid between $s$ and $s+\rmd s$. This slice of fluid colored in blue has a momentum per unit length $\bp$, is submitted to the force $-\bF$ by the body (red), and the pressures forces $\bF_p$ and $-\bF_p - \rmd \bF_p$. In addition a flux of momentum with velocity $-u$ goes through the slice. 
}
\label{fig:notations}
\end{center}
\end{figure}

We consider an elongated fish of length $L$ swimming in a fluid otherwise at rest (figure~\ref{fig:notations}).  
The position of the fish backbone is described by the vector $\boldsymbol{s}(s,t)$, with $s \in [0, L]$ the curvilinear coordinate from head to tail. We  assume planar motion, so that $\boldsymbol{s}$ remains in the $(x,z)$-plane. Each section of the fish is considered elliptic with a semi-major axis $b(s)$ aligned with the vertical $\haty$ and a semi-minor axis $a(s)$.

%%%%%%%%%%%%%%
%%%%%%%%%%%%%%

At each point of the backbone, the local reference frame is $(\hatt,\hatn, \haty)$, such that $\hatt = \boldsymbol{s}'$ and $\hatn= \haty \times \hatt$ are the tangent and normal unit vectors.  
The local velocity of the backbone is $\bv(s,t) = \bsd$ and its curvature $\kappa$ satisfies $\hatt'= \kappa \,\hatn$. Hereinafter primes refer to the partial derivative with respect to $s$ and dots with respect to $t$. The local velocity is further decomposed into tangential and normal components: $\bv = u\, \hatt + v\, \hatn$. 

In Lighthill's elongated-body theory \citep{Lighthill1971}, the goal is to determine the potential flow and the hydrodynamic forces exerted on an elongated fish in motion. 
We therefore consider that the body is elongated, assuming that (1) $b\ll L$. 
To derive Lighthill's expression of the reactive force, we make two further assumptions: 
(2) the cross-section and velocity vary slowly along $s$ ($b'\ll 1$, $bv'\ll v$); and 
(3) the curvature is small compared with $1/b$ ($\kappa b\ll 1$). 
The same assumptions are implicitly made in Lighthill's original derivation.

%%%%%%%%%%%%%%%%%%%%%%%%%%%%%%%%%%%%
\subsection{Derivation by momentum conservation}
\label{sec:momentum_derivation}

\cite{Lighthill1971} originally derived the reactive force using a control volume upstream of the plane normal to the backbone at the tail ($s=L$). Here, we will slightly adapt his derivation based on an argument of momentum conservation, using, as control volumes, slices of fluid located between $s$ and $s+\rmd s$ (figure \ref{fig:notations}b-c). 

Like \cite{Lighthill1971}, we consider the momentum $\bp$ (per unit $s$) attached to such a slice. If the local curvature $\kappa$ of the body is sufficiently small, then this momentum can be computed at leading order from that of a section of straight cylinder of the same cross-section: $\bp=m_a v\,\hatn$ with $m_a(s)=\pi b^2$ the section's added mass (see \S\ref{sec:straight}).  

Also, as for a straight cylinder, the fluid upstream of the slice exerts a  pressure force ${\bF_p=\frac{1}{2}m_a v^2\hatt}$, with $m_a$, $v$ and $\hatt$ evaluated at $s$ (see \S\ref{sec:straight}). The fluid downstream exerts an opposite force with quantities evaluated at $s +\rmd s$. The conservation of momentum for the slice of fluid $[s,\,s+\rmd s]$ can thus be written in the framework attached to the body
\begin{equation}
	\partial_t \bp=-\partial_s(-u\bp)-\partial_s \bF_p-\bF,
\end{equation}
where the three right-hand side terms correspond to the convective flux of momentum along the body (the fluid moves with a velocity $-u$ with respect to the body in the chosen framework), the pressure forces exerted by the upstream and downstream fluid and the net force applied by the body. 
Using the expressions for $\bp$ and $\bF_p$ given above, this leads to Lighthill's reactive force per unit length
\begin{equation}\label{eq:Lighthill_force}
\bF=-\partial_t(m_a v\,\hatn)+(m_a uv\,\hatn)'-\frac{1}{2}(m_a v^2\,\hatt)',
\end{equation}
exerted on the body's cross-section at $s$. 
Equivalently, using the kinematic relationship $\partial_t \hatn=-(\kappa u+v')\hatt$, this force can be written as
\begin{align}\label{eq:LAEBT}
\bF =-\left[m_a\dot{v}-(m_a uv)'+\frac{1}{2}m_a v^2\kappa \right]\hatn-\frac{1}{2}m_a'v^2\hatt,
\end{align}
which makes it apparent that Lighthill's force is mostly normal to the backbone, the only tangential term being linked to cross-section variations. Note that the local force is still normal to the surface, as expected for pressure forces.  
In \eqref{eq:LAEBT}, we also see that the first term corresponds to the acceleration of the fluid, the second term corresponds to the flux of momentum and is linear in $v$. However, the third term is non-linear in $v$ and involves the curvature. It is akin to a centrifugal acceleration and can be thought of as the necessary force to direct an added mass of water along a curved trajectory. This term cannot be found with a linear approach such as the one used in \cite{Lighthill1960}. 

\ce{
The expression of Lighthill's reactive force, given in \eqref{eq:Lighthill_force} and \eqref{eq:LAEBT}, is not new. It can be inferred from the arguments of \cite{Lighthill1971} and it has been rederived rigorously by \cite{Candelier2011} (equation 3.25 in their paper). 
In his original paper, \cite{Lighthill1971} applies momentum conservation to the full volume comprising the slices upstream of the trailing edge. His calculation of the total reactive force can be recovered by integrating \eqref{eq:Lighthill_force} between $s=0$ and $L$ 
\begin{equation}\label{eq:Lighthill_integral}
\mathbf{F}=\int_0^L \bF \,\rmd s = 
	\left[m_a uv\,\hatn-\frac{1}{2}m_a v^2\,\hatt\right]_{s=L}
	-\frac{\rmd}{\rmd t} \int_0^L m_a v\,\hatn \rmd s ,
\end{equation}
assuming that $m_a=0$ at $s=0$. Equation \eqref{eq:Lighthill_integral} is similar to equation (6) in \cite{Lighthill1971}. 
It can be further simplified by assuming a periodic motion and taking the time average of \eqref{eq:Lighthill_integral} projected along $\hatz$ to calculate the mean thrust as
\begin{equation}\label{eq:Lighthill_thrust}
\langle T \rangle = 
	\langle -\mathbf{F}\cdot\hatz \rangle = 
	\langle m_a v\left(\dot{x}-\tfrac{1}{2}vz'\right) \rangle_{s=L} ,
\end{equation}
this equation being similar to equation (7) in \cite{Lighthill1971}. It is worth mentioning that this expression of the reactive thrust is different from the result of slender-body theory, equation (7) in \cite{Lighthill1960}, which is obtained in the limit of small transversal displacements (i.e. $x'\ll 1$ and $\dot{x}\ll U$) and is expressed as $\langle T \rangle = \frac{1}{2} m_a \langle \dot{x}^2 - U^2 x'^2\rangle_{s=L}$ with the present notations, with $U$ the average swimming velocity. }

We have revisited Lighthill’s momentum conservation arguments to derive the reactive force, as expressed in \eqref{eq:Lighthill_force} and \eqref{eq:LAEBT}. While \cite{Candelier2011} rigorously confirmed the validity of this force, the momentum conservation approach may still raise concerns for some readers. The main issue lies in the implicit assumption that, at leading order, the local flow properties remain similar to those around a straight cylinder. However, both the momentum $\bp$ and the pressure force $\bF_p$ are influenced by the flow at large distances ($r \gtrsim L$), where the flow is affected by the body’s global motion, leading to significant deviations from the flow around a straight cylinder.

%%%%%%%%%%%%%%%%%%%%%%%%%%%%%%%%%%%%
\subsection{Derivation using a distribution of dipoles and sources}
\label{sec:dipole}

We will now derive Lighthill's force by reconstructing the flow potential with an approach similar to the one used for a straight cylinder in \S\ref{sec:straight}.
The differences are that the cylinder is now curved, its cross-section varies slowly along the backbone, and the velocity has a tangential component. 
This derivation will be valid in the limit of asymptotically small local curvature and rates of change of the cross-section radius and normal velocity. In the following, we will thus consider the distinguished scaling where $\kappa b$, $a'$, $b'$, and $v' b/u$ are small quantities of the same order $\varepsilon$.

By analogy with the case of a straight cylinder, we first consider a distribution of dipole singularities along the ribbon spanning between the section's foci, whose intensity $\bG(s, \eta) = P(s,\eta)\,\hatn$ is locally equal to that of a straight cylinder of identical cross-section, with
\begin{equation}\label{eq:dipole_density}
P(s, \eta) = \frac{2 b v}{b-a} \sqrt{c^2 - \eta^2},\qquad -c<\eta<c.
\end{equation}
where $a$, $b$, $c=\sqrt{b^2-a^2}$, and $v$ are functions of $s$. 
 
As we shall see below, this dipolar density is not sufficient to enforce the impermeability boundary condition on the cylinder's surface in $\bs(s) + \bS(s,\theta)$, with $\bS = a\cos\theta\,\hatn + b\sin\theta\,\haty$
\begin{equation}\label{eq:BCdipole}
\bnabla \phi \cdot \bN = \bv \cdot \bN = bv \cos\theta - u (a' b \cos^2 \theta + a b' \sin^2\theta),
\end{equation}
with  $\bN(s,\theta) = - u (a' b \cos \theta + a b' \sin\theta)\, \hatt + b\cos\theta\, \hatn + a\sin\theta \,\haty$, a (non-unitary) vector normal to the surface. 

We note with a tilde the velocity potential due to the distribution of dipoles given by \eqref{eq:dipole_density}. This potential $\phialt$ at a point $\bx$ in a plane normal to the backbone in $s=s_0$ is obtained by summing the contributions of the dipoles
\begin{equation}\label{eq:phi_curved}
\phialt(\bx)  =  \int_{0}^{L} \int_{-c}^{c} -\frac{\bG(s, \eta)\cdot\left(\bx-\bX(s,\eta)\right)}{4\pi\|\bx-\bX(s,\eta)\|^3}\rmd \eta\, \rmd s,
\end{equation}
with $\bX(s,\eta)$ a point on the ribbon.  When considering the flow near the cylinder surface, the above integral is dominated by local terms, i.e. $\st=s-s_0=O(b)$. As a consequence, %changing the origin of the labframe to $\bX(s_0)$ and the origin of $s$ to $s_0$, 
the integrand of \eqref{eq:phi_curved} can be computed by locally approximating the centerline with its tangent circle of curvature $\kappa_0=\kappa(s_0)$
\begin{subequations}\label{eq:Taylor_expansions}
\begin{eqnarray}
\bX(\st, \eta) & = & \eta\,\haty + \st\,\hatz + \frac{1-\cos (\kappa_0 \st)}{\kappa_0}\,\hatx + O\left(\varepsilon^2\right), \\
\bG(\st, \eta) & = & P_0(\eta) \,\hatx + P_0'(\eta) \st\,\hatx - P_0(\eta) \kappa_0 \st\,\hatz + O\left(\varepsilon^2\right),
\end{eqnarray}
\end{subequations}
where all quantities with a subscript 0 are evaluated at $s=s_0$, e.g., $P_0(\eta) = P(s_0,\eta)$ and $P_0'(\eta) = \displaystyle\frac{\partial P}{\partial s}(s_0,\eta)$, where $P$ is given by \eqref{eq:dipole_density} and the framework $(\hatx,\haty,\hatz)$ is centered onto $\bs(s_0)$ and aligned with the local framework $(\hatn_0,\haty,\hatt_0)$. 
Keeping the cosine with argument $(\kappa_0\st)$ (arising from the tangent circle approximation) instead of its leading order expansions for small $\st$ (tangent parabola approximation)  allows us to ensure convergent integrals in the following. 

Inserting the expansions of \eqref{eq:Taylor_expansions} into \eqref{eq:phi_curved} and integrating along $\st$ leads to ${\phialt =\phialt_0 + \phialt_1 + O(\varepsilon^2)}$, with
\begin{subequations}\label{eq:dipoles_only}
\begin{eqnarray}
\phialt_0 &  = & \int_{-c}^{c}-\frac{1}{2 \pi} 
	\frac{P_0(\eta)x}{x^2+(y-\eta)^2}  \rmd \eta ,\\
\phialt_1 &  = & \frac{\kappa_0}{2} \int_{-c}^{c}-\frac{1}{2 \pi}\left( 
		\frac{P_0(\eta)x^2}{x^2+(y-\eta)^2} - P_0(\eta)\log\sqrt{x^2+(y-\eta)^2}
		\right) \rmd \eta .
\end{eqnarray}
\end{subequations}
The $O(1)$ potential $\phialt_0$ is the same as the potential of an infinite straight cylinder given by \eqref{eq:phi_straight}. The $O(\varepsilon)$ potential $\phialt_1$ is proportional to the curvature $\kappa_0$ and can be divided into two terms: one equal to $\kappa_0 \phialt_0 x/2$ and the other corresponding to a distribution of sources with density $\kappa_0 P_0(\eta)/2$.

As mentioned above, this potential $\phialt$ does not satisfy the boundary condition \eqref{eq:BCdipole} since one can readily verify that on the surface  $\bS(\theta) = a\cos\theta \,\hatx + b\sin\theta\,\haty$ 
\begin{equation}\label{eq:BCwrong}
\bnabla \left(\phialt_0 +  \phialt_1\right)\cdot \bN = b v \cos\theta  + \frac{1}{2}\kappa a b v , 
\end{equation}
where we have dropped the subscripts $0$ since what is true at $s_0$ is also true for any $s$. 
The correct boundary conditions \eqref{eq:BCdipole} can be recovered if we add a distribution of sources/sinks of order $\varepsilon$ along the ribbon of the form $Q = Q_1+Q_2+Q_3$, with
\begin{subequations}\label{eq:sourcesQ}
\begin{eqnarray}
Q_1(s,\eta) &=& -v \frac{\kappa a b}{\sqrt{c^2 - \eta^2}} ,\\
Q_2(s,\eta) &=& -u \frac{ab'+a'b'}{\sqrt{c^2 - \eta^2}},\\
Q_3(s,\eta) &=& -u \frac{ab' - a'b}{(b-a)^2}\frac{c^2 - 2\eta^2}{\sqrt{c^2 - \eta^2}}.
\end{eqnarray}
\end{subequations}
In Appendix \ref{sec.appendix3}, we show how these source/sink terms are calculated using the framework of complex potentials in the $(x,y)$-plane. 
The term $Q_1$ is linked to the change of volume of a body with curvature that is moving with a velocity normal to its centerline. For a portion of centerline $\rmd s$ with volume $\pi ab \rmd s$, after an infinitesimal time $\rmd t$, the centerline will have shrunk/expanded, such that the volume is now $\pi ab \rmd s (1 - \kappa  v \rmd t)$, hence the need of a source/sink term per unit length equal to $\int_{-c}^{c}Q_1\rmd\eta=- \pi ab \kappa  v$. The source $Q_2$, which sums up to $\int_{-c}^{c}Q_2\rmd\eta=-\pi u (ab)'$, is proportional to the change of volume of the cross-section and corresponds to an inward (resp. outward) flow when the section surface is decreasing (resp. increasing) with $s$.  Finally, the source $Q_3$, which sums up to zero, corresponds to a quadripolar flow associated with a rate of change of the eccentricity without a change of the elliptical section surface.  

The potential is given by
\begin{equation}\label{eq:phi_curved_corr}
\phi(\bx) 	= \int_{0}^{L} \int_{-c}^{c}  -\frac{1}{4\pi} \left( 
	\frac{\bG(s,\eta)\cdot(\bx-\bX(s,\eta))}{\|\bx-\bX(s,\eta)\|^3} + 
	\frac{Q(s,\eta)}{\|\bx-\bX(s,\eta)\|} 
\right)\rmd \eta\, \rmd s,
\end{equation}
with the dipole density $\bG$ given by \eqref{eq:dipole_density} and the source density $Q$ given by (\ref{eq:sourcesQ}a--c).

From this potential $\phi$, the pressure $p$ near the cylinder's surface can be obtained through the generalised Bernoulli equation \eqref{eq:p_straight}.
By integrating the pressure forces, one can then compute the force per unit length exerted by the fluid on the cylinder 
\begin{equation}\label{lighthill_f2}
\bF = \int_{-\pi}^{\pi} - p\, \bN S(\theta)\,\rmd \theta 
    = \left(m'_a uv + m_a u v' + \frac{1}{2}m_a v^2\kappa -m_a\dot{v} \right)\hatn-\frac{1}{2}m_a'v^2\,\hatt,
\end{equation}
where $S(\theta) = (1-\kappa b \cos\theta)$ accounts for the excess of surface due to curvature (more details on this calculation are given in appendix \ref{sec.appendix4}). Using the identity $v'=-\kappa u$, we note that this equation is similar to \eqref{eq:LAEBT}. We have thus recovered  Lighthill's force with the dipole density $\bG(s,\eta)$ given by \eqref{eq:dipole_density} and the source density $Q(s,\eta)$ given by (\ref{eq:sourcesQ}a--c) along a ribbon spanning between the foci of the ellipse. 

The present derivation \ce{of Lighthill's reactive force} has the additional benefit of providing an explicit computation of the flow field. \ce{This was not directly possible with Lighthill's original formulation \citep{Lighthill1971}, but \cite{Mougel2016} have shown how the potential jump could be related to the flow field through the use of Green's second identity in a flapping rectangular plate. Here, we have provided a simple way to compute the flow for an arbitrary varying elliptic cross-section. It will} allow us to compare the predictions of elongated-body theory to full numerical simulations.

%%%%%%%%%%%%%%%%%%%%%%%%%%%%%%%%%%%%%%%%%%%%%%%%%%%%%%%%%%%%%%%%%%%%%%%
\section{Comparison with numerical simulations}
\label{sec.Comparison_numerics}

%From a practical point of view, it is useful to change $z$ into $c(s) \zeta$ in the integral above such that the $(\zeta, s)$ domain of integration becomes a rectangle. In addition, if one is interested in the flow in the plane $z=0$, the $\zeta$ integrations of $\bG$ and $Q$ can be performed analytically. It means that the flow in the plane $z=0$ can be computed from a numerically-evaluated linear integrals along $s$. 

In \cite{Candelier2011}, the authors compared the reactive force predicted by Lighthill's elongated-body theory with the results of a numerical simulation performed with a RANS (Reynolds averaged Navier--Stokes) solver. They show a good agreement for all components except for the longitudinal one, which is anticipated since viscous effects play an important role for this component. 

Here, to illustrate the flow predicted by Lighthill's elongated-body theory, we will use the potential given in \eqref{eq:phi_curved_corr}. 
\ce{We will further consider the distribution of dipoles at $s=L$ is shed in a wake by virtue of Kelvin's circulation theorem. We will assume that these dipoles do not diffuse and are not advected by the flow created by the fish's motion or by themselves, thus creating a ``frozen wake'' along the trajectory of the fish's tail that contributes to the total flow (but not to the reactive force).} 
This creates a periodic ribbon of dipoles, along a sinusoidal path with the same amplitude as the tail beat amplitude and wavelength $U f$, with $U$ the swimming velocity and $f$ the tail beat frequency. Note that sources/sinks cannot be shed in the wake because the flow has to remain incompressible. 

\ce{The periodic ribbon of dipoles in the wake with dipolar intensity $P(s,\eta)$ is equivalent to a ribbon of tangential velocity jump, or equivalently a distribution of vorticity $\boldsymbol{\gamma}$. In particular, the $y$-component of the wake vorticity in the midplane ($\eta=0$) is equal to
\begin{equation}\label{wake_vorticity}
\boldsymbol{\gamma}\cdot\haty = \left(\frac{\partial P}{\partial s}\right)_{\eta=0} = - \frac{2b(L) \dot{v}}{\sqrt{\dot{x}^2 + \dot{z}^2}},
\end{equation}
where we have parametrised the wake by the time $t<0$ at which the dipoles have been shed at the position $\bs(L,t)$, with $a(L)=0$, $P(t,\eta=0) = 2b(L)v(t)$ and $\partial / \partial s = - \partial / \partial t (1 / \sqrt{\dot{x}^2 + \dot{z}^2})$ in the wake. 	}

This potential flow field is compared to the inviscid numerical simulations performed by \cite{Borazjani2008,Borazjani2009} on a model mackerel and a model lamprey. The shape and kinematics of the mackerel are taken from the experimental study of \cite{Videler1984}. For the lamprey, the shape is taken from a body scan performed by Frank Fish and the kinematics is inspired by a study on eels \cite{Tytell2004}. We used the same swimming kinematics and the same body shapes, only slightly modified to make the sections elliptical. 

\begin{figure}
\begin{center}
\includegraphics[width=0.99\textwidth]{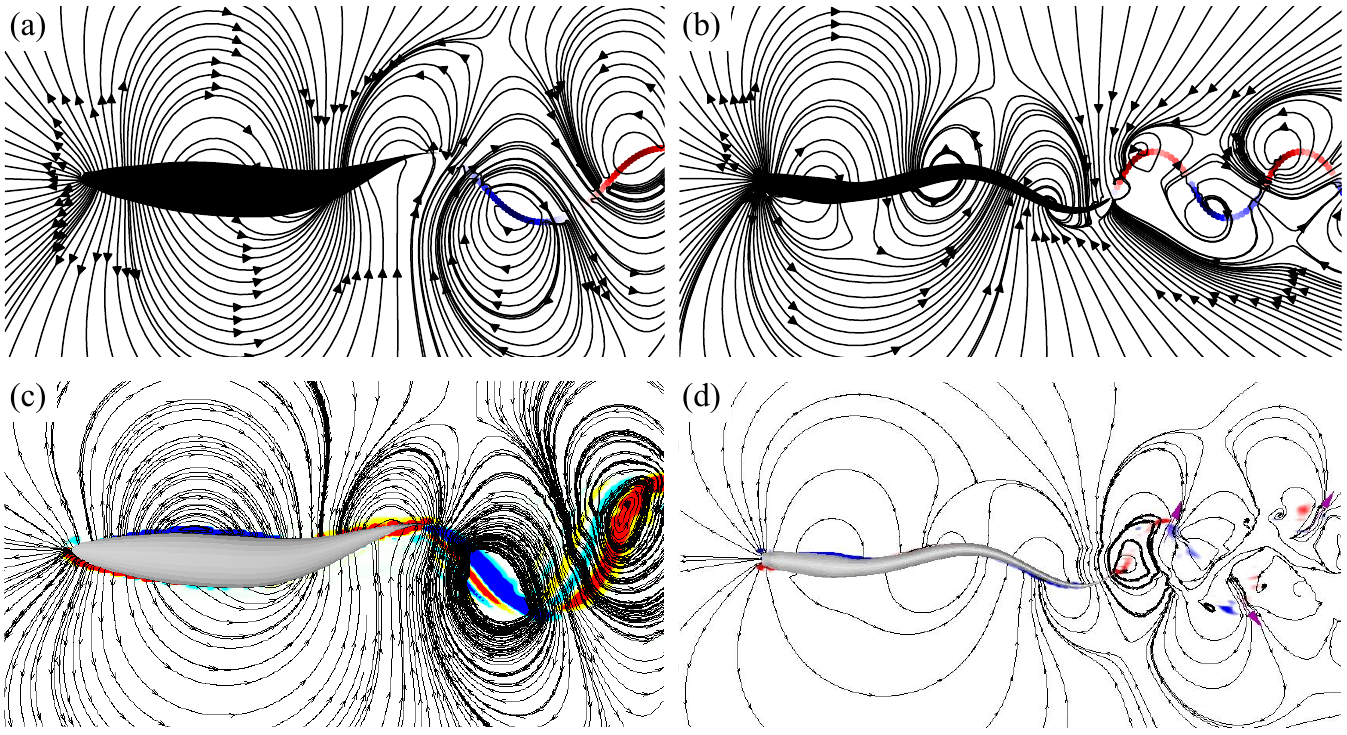}
\caption{Streamlines around a swimming mackerel (a, c) and a lamprey (b, d). In (a, b), the streamlines are computed from the dipole and source densities given by \eqref{eq:dipole_density} and (\ref{eq:sourcesQ}a--c). The coloured dots in the wake show the vorticity \ce{shed given by \eqref{wake_vorticity}}. In (c, d), numerical simulations of \cite{Borazjani2008} and \citep{Borazjani2009} for a potential flow. 
}
\label{fig:comparison}
\end{center}
\end{figure}

The comparisons of the streamlines in the horizontal mid-plane are shown in figure~\ref{fig:comparison}. Although the streamlines differ in their details, they look similar, showing that Lighthill's elongated-body theory is a good alternative to numerical simulations with the additional benefit of being computationally efficient and interpretable. 

%%%%%%%%%%%%%%
%%%%%%%%%%%%%%

\begin{figure}
\begin{center}
\includegraphics[width=0.99\textwidth]{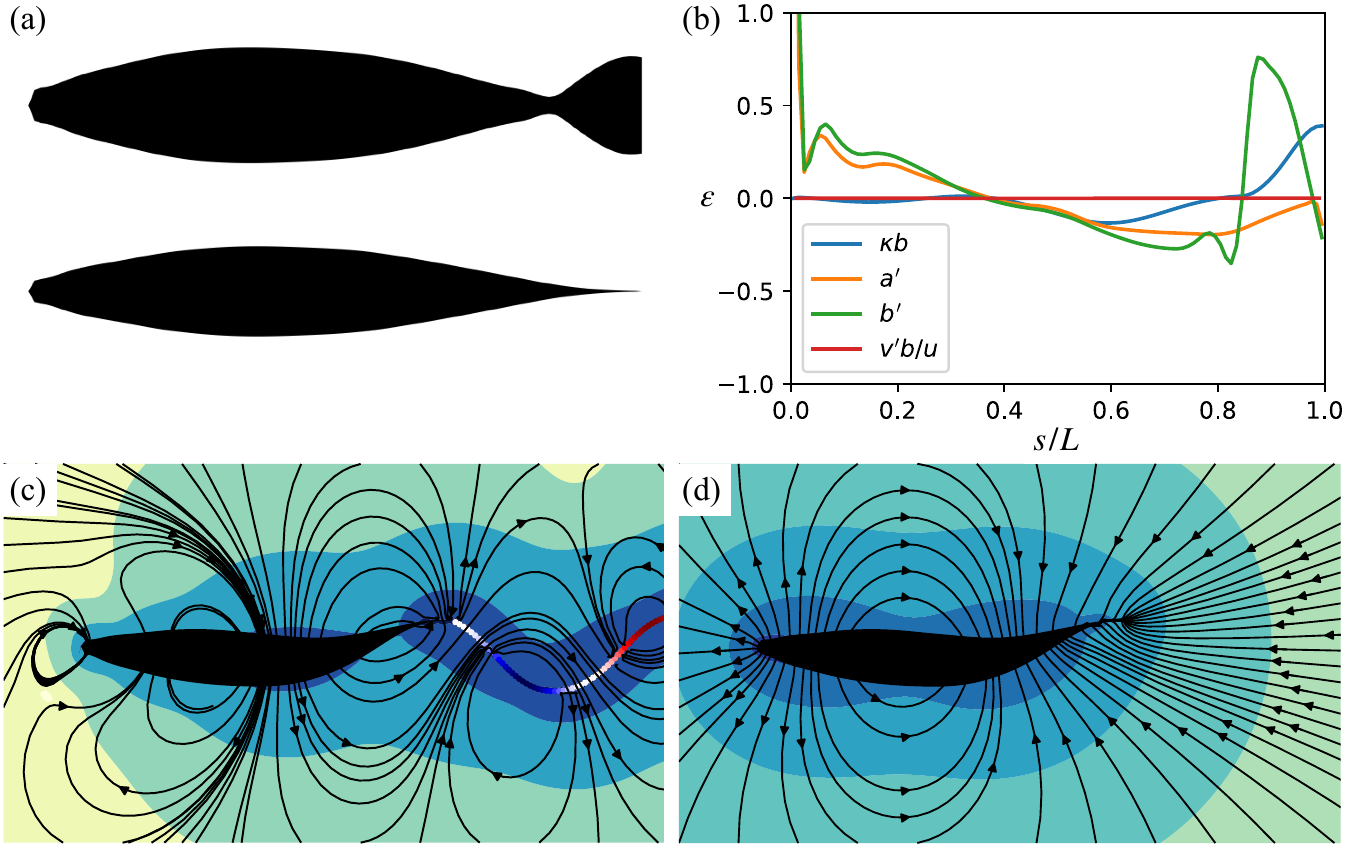}
\caption{
\ce{
(a) Side view and top view of the mackerel shape used to calculate the potential flow in figure~\ref{fig:comparison}a \citep{Borazjani2008}.
(b) The quantities $\kappa b$, $a'$, $b'$, and $bv'/u$, assumed to be small in Lighthill's elongated-body theory, are plotted as a function of the normalised curvilinear coordinate $s/L$.
(c) Contributions to the flow field of the dipole distribution $P(s,\eta)$ in the body and in the wake. 
(d) Same as (c) for the contributions of the source density $Q(s,\eta)$. The background colours show levels of the flow speed with the same scale.
} 
}
\label{fig:mackerel}
\end{center}
\end{figure}

\ce{To assess if the hypotheses of Lighthill's elongated-body theory hold in a realistic case, we have plotted together with the mackerel shape, the different quantities that are assumed of order $\varepsilon \ll 1$ (figure~\ref{fig:mackerel}a,b).
It shows that these quantities reach values close to 0.5 near the tail. Despite these large values, the streamlines predicted are close to the results of an Euler numerical simulations as shown by comparing figures \ref{fig:comparison}a and \ref{fig:comparison}c \citep{Borazjani2008}.
To quantify the importance of the different contributions to the flow field, we have plotted, in figure \ref{fig:mackerel}c, the flow due to the leading order contribution (the dipole density $P$ in the body and in the wake) and, in figure \ref{fig:mackerel}d, the flow due to the order $\varepsilon$ contribution (the source density $Q$).
Surprisingly, the flow due to the source density $Q$ is not merely a correction to the main flow, but is of the same or even larger magnitude, especially around the body. It shows the importance of this contribution to the flow that amounts, in the far field, to a dipolar flow field (the head acting as a source and the tail as a sink).
}

%%%%%%%%%%%%%%
%%%%%%%%%%%%%%

%%%%%%%%%%%%%%%%%%%%%%%%%%%%%%%%%%%%%%%%%%%%%%%%%%%%%%%%%%%%%%%%%%%%%%%
\section{Discussion}
\label{sec.Discussion}

In this paper, we revisited Lighthill’s elongated-body theory and introduced a new derivation of the reactive hydrodynamic force acting on an undulating swimming fish. We extended Lighthill's original framework by making use of the essential singularities of the Laplace equation. This approach allowed us not only to confirm the validity of Lighthill’s force predictions but also to compute the full three-dimensional flow field generated by the swimming motion. Our results align well with numerical simulations of realistic fish geometries, offering a deeper insight into the hydrodynamic forces and flows involved in fish locomotion.

\ce{The mean thrust obtained by \cite{Lighthill1971} and recovered here in \eqref{eq:Lighthill_thrust} only depends on the kinematics of the tail's trailing edge. As already pointed out by \cite{Lighthill1960,Lighthill1971} and \cite{Pedley1999}, this remarkable result should not hide the fact that a swimming fish cannot fully prescribe its kinematics because of recoil motions.
The recoil consists of additional time-dependent rigid body translation and rotation that must be added to the prescribed kinematics such that the fish obeys Newton's second law of motion.}

Despite its usefulness, Lighthill’s elongated-body theory has two notable limitations. The first is its reliance on potential flow, which inherently neglects viscous effects. This limitation can be addressed by incorporating viscous forces, as suggested by \cite{Candelier2011}, through an approach similar to the Morison equation for vibrating cylinders \citep{sarpkaya1986force}.

The second limitation is that Lighthill’s assumptions violate the unsteady Kutta condition at the caudal fin’s trailing edge \citep{Crighton1985}: \ce{there indeed exists a finite pressure jump, whereas the Kutta condition stipulates that this pressure jump should be zero. To address this limitation, one possibility is to calculate explicitly the three-dimensional potential flow near the trailing edge, for instance, using a vortex panel method \citep{Cheng1991,Cheng1998}. The main effect of the Kutta condition is a reduction of the reactive force near the trailing edge as compared to Lighthill's approach, which implicitly assumes a `frozen wake' as done in the present paper. \cite{Yu2018} recently proposed a simple method to incorporate this effect, reconciling the elongated-body theory with the Kutta condition.}

Lighthill’s elongated-body theory continues to offer potential for future works. Its computational efficiency and ability to handle non-linear deformations make it particularly well-suited for optimization studies, especially when complemented by viscous force corrections \citep{Candelier2011}, as demonstrated in recent works \citep{eloy2012optimal, eloy2013best}. Furthermore, the extension introduced by \cite{Candelier2013}, which incorporates background flow, opens up new opportunities to explore collective dynamics in fish schools \citep{weihs1975some, filella2018hydrodynamic, Li2019}. With these refinements, Lighthill’s theory remains a valuable tool for advancing our understanding of fish locomotion and hydrodynamic interactions.

%%%%%%%%%%%%%%%%%%%%%%%%%%%%%%%%%%%%%%%%%%%%%%%%%%%%%%%%%%%%%%%%%%%%%%
\begin{acknowledgments}

%\textbf{Supplementary data.} Supplementary material and movies are available at https://doi.org/10.1017/jfm.2019...

\textbf{Acknowledgements.} A preliminary version of this work involved contributions from Frédéric Boyer, Fabien Candelier, and Jérôme Mougel, to whom we extend our warm thanks for their stimulating discussions and encouragement. 
We are grateful to Iman Borazjani and Fotis Sotiropoulos for sharing their simulation results. 

\textbf{Funding.} This work was
supported by the European Research Council (ERC) under the European Union’s Horizon 2020
research and innovation program (Grant Agreement No. 714027 to S.M.).

\textbf{Declaration of interests.} The authors report no conflict of interest.

%\textbf{Data availability statement.} The data that support the findings of this study are openly available in [repository name] at http://doi.org/[doi], reference number [reference number].

\textbf{Author ORCID.} %Authors may include the ORCID identifers as follows. F. Smith, https://orcid.org/0000-328 0001-2345-6789; B. Jones, https://orcid.org/0000-0009-8765-4321
C. Eloy, https://orcid.org/0000-0003-4114-7263;
S. Michelin, https://orcid.org/0000-0002-9037-7498 

%\textbf{Author contributions.} Authors may include details of the contributions made by each author to the manuscript, for example, “A.G. and T.F. derived the theory and T.F. and T.D. performed the simulations. All authors contributed equally to analysing data and reaching conclusions, and in writing the paper.”

\end{acknowledgments}

\appendix

%%%%%%%%%%%%%%%%%%%%%%%%%%%%%%%%%%%%%%%%%%%%%%%%%%%%%%%%%%%%%%%%%%%%%%%
\section{Complex representation and boundary condition for a straight cylinder}
\label{sec.appendix1}

In the bidimensional limit, valid for $r \ll L$ and far from the cylinder's ends, the velocity potential is given by \eqref{eq:phi_straight} with $g\approx 1$. 
To evaluate the induced velocity and pressure near the cylinder's surface, it is convenient to use the complex variable $\zeta = x+\rmi y$ and the complex potential $f= \phi + \rmi \psi$ in the $\zeta$-plane, with $\psi$ the streamfunction. 
The complex potential $f$ of a density of dipole $P$ distributed between $\zeta = -\rmi c$ and $\rmi c$ is given by
\begin{equation}\label{eq:complex_f}
f(\zeta) 
		 = \int_{-c}^{c} - \frac{P(\eta)\rmd \eta}{2 \pi (\zeta -\rmi\eta)} 
		 = \frac{bv}{b-a} \left(\zeta - \sqrt{c^2+\zeta^2} \right) .
\end{equation}
It is easy to verify that the boundary condition \eqref{eq:BC_straight} is satisfied: within this complex formulation, it can be written in $\zeta = a \cos \theta + \rmi b \sin \theta$ (on the cylinder's surface) as
\begin{equation}\label{eq:BC_complex}
\Real \left(w N \right) = \Real (v N) = bv \cos\theta, 
\end{equation}
with $w=\rmd f/\rmd \zeta$ the complex velocity and $N = b \cos \theta + \rmi a \sin \theta $, a (non-unitary) vector normal to the ellipse. 

%%%%%%%%%%%%%%%%%%%%%%%%%%%%%%%%%%%%%%%%%%%%%%%%%%%%%%%%%%%%%%%%%%%%%%%
\section{Integration of pressure forces for a straight cylinder}
\label{sec.appendix2}

The pressure field can be obtained from the generalised Bernoulli equation \eqref{eq:p_straight}. In the case of a straight cylinder, the only contributing term to the total pressure force is the $x$-odd component proportional to $\dot{\phi}$. Using \eqref{eq:complex_f}, the meaningful pressure term can therefore be evaluated along the ellipse contour $\zeta = a \cos \theta + \rmi b \sin \theta$ as
\begin{equation}
p(\zeta) = - \rho \dot{\phi} 	= - \rho \Real (\dot{f}) = \rho b \dot{v} \cos\theta,
\end{equation}

When integrated along the ellipse contour, this pressure distribution gives a force per unit length along $x$
\begin{equation}
f_a = \int_{-\pi}^{\pi} - p(\zeta)N\,\rmd \theta 
	= -\rho  b^2 \dot{v} \int_{-\pi}^{\pi} \cos^2\theta\,\rmd \theta 
    = -m_a\dot{v},
\end{equation}
with $m_a=\rho\pi b^2$ the added mass per unit length of the cylinder.

%%%%%%%%%%%%%%%%%%%%%%%%%%%%%%%%%%%%%%%%%%%%%%%%%%%%%%%%%%%%%%%%%%%%%%%
\section{Calculation of source/sink terms in the general case}
\label{sec.appendix3}

From the Taylor expansion of the integrand appearing in \eqref{eq:phi_curved}, the flow potential was obtained as $\phialt = \phialt_0 + \phialt_1 + O(\varepsilon^2)$, with $\phialt_0$ and $\phialt_1$ given by \eqref{eq:dipoles_only}. From these expressions,  the flow at the cylinder's surface can be expressed using the complex potential formulation.

The leading order potential $\phialt_0$ is associated with the flow $w_0 = \rmd f_0 / \rmd \zeta$, with $f_0$ given by \eqref{eq:complex_f}. The $O(\varepsilon)$-potential $\phialt_1$ can be decomposed into two terms: $\phialt_{1,0} = \kappa \phialt_0 x /2$ associated with the flow $w_{1,0}$ and $\phialt_{1,1}$ associated with a distribution of sources of density $\kappa P(\eta) /2$ with complex velocity $w_{1,1}$. At the cylinder's surface $\zeta = a \cos \theta + \rmi b \sin \theta$, the complex velocities $w_{1,0}$ and $w_{1,1}$ can be written
\begin{subequations}\label{w_10}
\begin{eqnarray}
w_{1,0} &  = & \frac{\kappa}{2} \left( a\cos\theta w_0 + \Real(f)\right),\\
w_{1,1} &  = & \frac{\kappa}{2} \frac{bv}{b-a} \left(\sqrt{c^2+\zeta^2} - \zeta \right) .
\end{eqnarray}
\end{subequations}

Using this, one can show that the flow at the cylinder's surface due to the potential $\phialt$ is such that
\begin{equation}\label{spurious_flow}
\Real \left((w_0 + w_{1,0} +w_{1,1}) N \right) = bv \cos\theta + \frac{\kappa}{2} abv.  
\end{equation}
What we would like instead is the boundary condition \eqref{eq:BCdipole}, which would give with the complex representation the following boundary condition in $\zeta = a \cos \theta + \rmi b \sin \theta$
\begin{equation}\label{BC_Taylor}
\Real \left(w N \right) = bv \cos\theta - u \left(a'b\cos^2\theta + a b' \sin^2\theta\right).  
\end{equation}

As we can see from the difference between  \eqref{BC_Taylor} and \eqref{spurious_flow}, the dipole density is not enough to satisfy the correct boundary condition because we are missing a source/sink flow. 

A distribution of sources between the foci of the elliptic cross section ($-c<y<c$) with density $Q=2q / \sqrt{c^2 - y^2}$ [resp. $Q=2q \sqrt{c^2 - y^2}$] produces a flow with a complex velocity $w = q/\sqrt{c^2 + \zeta^2}$ [resp. $w=q(\sqrt{c^2 + \zeta^2} - \zeta)$]. 
This flow is associated with the boundary condition at the surface $\zeta=a\cos\theta+\rmi b\sin\theta$ that writes $\Real(w N) = q$ [resp. $\Real(w N)=q (b-a) (a\sin^2\theta + b\cos^2\theta)$].  
By combining these two types of sources, we can ensure the correct boundary condition \eqref{BC_Taylor} by adding the sources $Q_1$, $Q_2$, and $Q_3$ given in (\ref{eq:sourcesQ}a--c).

%%%%%%%%%%%%%%%%%%%%%%%%%%%%%%%%%%%%%%%%%%%%%%%%%%%%%%%%%%%%%%%%%%%%%%%
\section{Integration of pressure forces in the general case}
\label{sec.appendix4}

We now want to compute the pressure at the body's surface by using the generalised Bernoulli equation \eqref{eq:p_straight}.
To calculate this pressure, the idea is to use the Taylor expansions (\ref{eq:Taylor_expansions}a-b) and keep only the terms up to order $\varepsilon$. 
This yields the pressure in the $\zeta$-plane
\begin{multline}\label{p_complex}
\frac{p - p_\infty}{\rho} =  
	- \frac{1}{2}\left( |w_0|^2 + w_0\left( \bar{w}_{1,0} + \bar{w}_2\right) + \bar{w}_0\left( w_{1,0} + w_2 \right)\right)
	\\ +\Real \left[
		u f'_0 + v\left(w_0 + w_{1,0} + w_2\right)
		- \dot{f_0} 
	 \right],
\end{multline}
with the overbar denoting complex conjugation, $w_{1,0}$ given by (\ref{w_10}a), $w_2 = Q_2 (\sqrt{c^2 + \zeta^2} - \zeta)/2$, $Q_2$ given by (\ref{eq:sourcesQ}b), and $f'_0 = \rmd f_0 / \rmd s$. For simplicity, we have omitted the terms corresponding to source terms with densities proportional to $1/ \sqrt{c^2 - y^2}$ such as $Q_1$, as they do not contribute to the pressure. 

From \eqref{p_complex} and \eqref{lighthill_f2} we can compute the longitudinal and normal components of the pressure force, which, after a lengthy but straightforward calculation, simplifies into
\begin{subequations}
\begin{eqnarray}
f_t &  = & \int_{-\pi}^{\pi} -p (a'b\cos^2\theta - ab'\sin^2\theta) \,\rmd \theta = -\frac{1}{2}m'_a v^2,\\
f_n &  = & \int_{-\pi}^{\pi} -p b\cos\theta(1-\kappa a\cos\theta)\,\rmd \theta = m'_a uv + m_a u v' + \frac{1}{2}m_a v^2\kappa -m_a\dot{v}.
\end{eqnarray}
\end{subequations}
The above expressions correspond to the result given in \eqref{lighthill_f2}. 

\bibliographystyle{jfm}
\bibliography{./biblio_Lighthill}

\end{document}